\newif\iffigs\figstrue
\documentclass[12pt]{article}
\usepackage{latexsym,amssymb,lscape}
\usepackage{graphicx}        % pacchetto grafico standard LaTeX
\newtheorem{congettura}{Conjecture}[section]

\def\IC{\relax\,\hbox{$\inbar\kern-.3em{\rm C}$}}
\def\IG{\relax\,\hbox{$\inbar\kern-.3em{\rm G}$}}
\def\IB{\relax{\rm I\kern-.18em B}}
\def\ID{\relax{\rm I\kern-.18em D}}
\def\IL{\relax{\rm I\kern-.18em L}}
\def\IF{\relax{\rm I\kern-.18em F}}
\def\IH{\relax{\rm I\kern-.18em H}}
\def\II{\relax{\rm I\kern-.17em I}}
\def\IN{\relax{\rm I\kern-.18em N}}
\def\IP{\relax{\rm I\kern-.18em P}}
\def\IQ{\relax\,\hbox{$\inbar\kern-.3em{\rm Q}$}}
\def\bfzero{\relax\,\hbox{$\inbar\kern-.3em{\rm 0}$}}
\def\IK{\relax{\rm I\kern-.18em K}}
\def\IG{\relax\,\hbox{$\inbar\kern-.3em{\rm G}$}}
 \font\cmss=cmss10 \font\cmsss=cmss10 at 7pt
\def\IR{\relax{\rm I\kern-.18em R}}
\def\ZZ{\relax\ifmmode\mathchoice
{\hbox{\cmss Z\kern-.4em Z}}{\hbox{\cmss Z\kern-.4em Z}}
{\lower.9pt\hbox{\cmsss Z\kern-.4em Z}} {\lower1.2pt\hbox{\cmsss
Z\kern-.4em Z}}\else{\cmss Z\kern-.4em Z}\fi}
\def\bfone{\relax{\rm 1\kern-.35em 1}}

\def\diag{{\rm diag}}

\def\inbar{\vrule height1.5ex width.4pt depth0pt}
\def\bfzero{\relax{\rm I\kern-.18em 0}}
\def\bfone{\relax{\rm 1\kern-.35em 1}}

\def\1bar{1\hskip -.275cm -}
\def\2bar{2\hskip -.275cm -}
\def\3bar{3\hskip -.275cm -}

\newsavebox{\uuunit}
\sbox{\uuunit}
                 {\setlength{\unitlength}{0.825em}
                      \begin{picture}(0.6,0.7)
                                      \thinlines
                                      \put(0,0){\line(1,0){0.5}}
                                      \put(0.15,0){\line(0,1){0.7}}
                                      \put(0.35,0){\line(0,1){0.8}}
                                     \multiput(0.3,0.8)(-0.04,-0.02){10}{\rule{0.5pt}{0.5pt}}
                      \end {picture}}

\makeatletter \@addtoreset{equation}{section} \makeatother
\newcommand{\be}{\begin{equation}}
\newcommand{\ee}{\end{equation}}
\newcommand{\ba}{\begin{eqnarray}}
\newcommand{\ea}{\end{eqnarray}}
\def\bfone{\relax{\rm 1\kern-.35em 1}}

\def\bfone{\relax{\rm 1\kern-.35em 1}}
\font\cmss=cmss10 \font\cmsss=cmss10 at 7pt
\newcommand{\nn}{\nonumber}
\newcommand{\so}{\mathfrak{so}}

\newcommand{\slal}{\mathfrak{sl}}

\begin{document}
\begin{titlepage}
%\begin{flushright}
%DFTT07/23\\
%JINR-E2-2007-145
%\end{flushright}
\vskip 0.2cm
\begin{center}
{\Large {\bf The Integration Algorithm \\
for Nilpotent Orbits of $\mathrm{G}/\mathrm{H}^\star$ Lax systems:
}}\\
{\Large {\bf \textit{i.e.} for Extremal Black Holes$^\dagger$
}}\\[0.3cm]
{\large Pietro Fr\'e$^{\tt a}$  and Alexander S. Sorin$^{\tt b}$}
{}~\\
\quad \\
{{\em $^{\tt a}$ Italian Embassy in the Russian Federation, \\
Denezhny Pereulok, 5, 121002 Moscow, Russia\\
{\tt  pietro.fre@esteri.it}\\
{\tt and}
\\
Dipartimento di Fisica Teorica, Universit\'a di Torino,}}
\\
{{\em $\&$ INFN - Sezione di Torino}}\\
{\em via P. Giuria 1, I-10125 Torino, Italy}~\quad\\
{\tt   fre@to.infn.it}
{}~\\
\quad \\
{{\em $^{\tt b}$ Bogoliubov Laboratory of Theoretical Physics,}}\\
{{\em Joint Institute for Nuclear Research,}}\\
{\em 141980 Dubna, Moscow Region, Russia}~\quad\\
{\tt sorin@theor.jinr.ru}
\quad \\
\end{center}
~{}
\begin{abstract}
Hereby we complete the proof of integrability of the Lax systems,
based on pseudo-Riemannian coset manifolds
$\mathrm{G}/\mathrm{H}^\star$, we recently presented in a previous
paper [arXiv:0903.2559]. Supergravity spherically symmetric black
hole solutions have been shown to correspond to geodesics in such
manifolds and, in our previous paper, we presented the proof of
Liouville integrability of such differential systems, their
integration algorithm and we also discussed the orbit structure of
their moduli space in terms of  conserved hamiltonians. There is a
singular cuspidal locus in this moduli space which needs a
separate construction. This locus contains the orbits of Nilpotent
Lax operators corresponding to extremal Black Holes. Here we
intrinsically characterize such a locus in terms of the
hamiltonians and we present the complete integration algorithm for
the Nilpotent Lax operators. The algorithm is finite, requires no
limit procedure and it is solely defined  in terms of the initial
data. For the $\mathrm{SL(3,\mathbb{R})/SO(1,2)}$ coset we give an
exhaustive classification of all orbits, regular and singular, so
providing general solutions for this case. Finally we show that
our integration algorithm  can be generalized to generic
non-diagonalizable (in particular nilpotent) Lax matrices not
necessarily associated with symmetric spaces.
\end{abstract}
\hrule {\footnotesize $^
\dagger $ This work is supported in part  by the Italian Ministry
of University (MIUR) under contracts PRIN 2007-024045. Furthermore
the work of A.S. was partially supported by the RFBR Grants No.
09-02-12417-$\mathrm{ofi\_m}$ , 09-02-00725-a, 09-02-91349-$\mathrm{NNIO\_a}$;
DFG grant No 436 RUS/113/669, and the Heisenberg-Landau
Program.}
\end{titlepage}
%\tableofcontents
\section{Introduction}
\label{intro} As announced in the abstract, this paper completes
the results of our very recent paper \cite{noiultimo} where we
discussed the classification of orbits for $\mathrm{G/H}^\star$
Lax systems, using the conserved hamiltonians arising from the
underlying Poisson structure and adapted Kodama's integration
algorithm \cite{kodama1} to this case. For the motivations, the
general setup and a more extended bibliography we refer the reader
to that publication. Assuming the framework of \cite{noiultimo} we
just focus on the problem of nilpotent orbits, that have a
distinct physical interest since they correspond to extremal Black
Holes. The authors of \cite{marioetal}, in parallel to our paper
\cite{noiultimo}, have also independently adapted Kodama
integration algorithm to the treatment of $\mathrm{G/H}^\star$ for
the case of non nilpotent initial conditions and have presented
some particular solutions.
\par
The results presented here are three:
\begin{description}
  \item[A.] The classification of exceptional orbits of Nilpotent Lax operators.
  \item[B.] The explicit integration algorithm for such operators.
  \item[C.] A generalized formula giving solely in terms of the initial data $L_0$
the integration of Lax equation for arbitrary non-diagonalizable
and diagonalizable Lax matrices $L$, not necessarily representing
symmetric spaces. This is specially useful if $L_0$ is
non-diagonalizable (in particular nilpotent) since this is the
only available integration formula.
  This result is presented in the appendix.
\end{description}
Although the presented integration algorithm is general, to illustrate it, we heavily rely on the simple
$\mathrm{SL(3,\mathbb{R})/SO(1,2)}$ example, already employed in \cite{noiultimo}.
The same paradigmatic example is utilized to explain the classification of orbits.
\section{The singular orbits of Nilpotent operators}
The integration algorithm determining the solution of the first order differential equations for the
tangent vector to a geodesic in a manifold $\mathrm{G}/\mathrm{H}^\star$ was discussed in \cite{noiultimo}
where their Liouville integrability was proved through the construction of the preserved hamiltonians, constructed
with the following procedure. Given the Lax operator $L$, fulfilling the $\eta$-symmetry condition
\begin{eqnarray}
(L~\eta)^{T} = L~\eta
 \label{LaxProp}
\end{eqnarray}
with
\begin{eqnarray}
\eta = \diag\left(-1,+1,...,-1,+1,+1,...+1\right)~, \quad p \leq q \quad ; \quad p+q \, = \, \mathrm{N}
 \label{metric}
\end{eqnarray}
the complete set of $p_\mathrm{N}$ functions $\mathfrak{h}_\alpha $ that are involutive with
respect to the Lie--Poisson bracket defined on the solvable Borel Lie algebra of $\mathrm{N \times N}$ upper triangular matrices is enumerated by an ordered pair of indices
\begin{equation}\label{orderedpair}
    \alpha \, = \, (a,b)
\end{equation}
where:
\begin{eqnarray}
  a &=&  0,...,\left[\frac{\mathrm{N}}{2}\right]~,\nonumber\\
  b &=& 1,..., \mathrm{N}-2a ~.
\end{eqnarray}
The functions $\mathfrak{h}_{ab}$
can be iteratively derived from the following relation:
\begin{eqnarray}
\label{tuttehamile} &&\det\left\{(L-\lambda)_{ij}:~a+1\leq i \leq
\mathrm{N}, ~ 1\leq j \leq
\mathrm{N}-a\right\}\nonumber\\&&=\mathcal{E}_{a 0
}\left(\lambda^{\mathrm{N}-2a} +\sum_{b=1}^{\mathrm{N}-2a}
\mathfrak{h}_{a b}~\lambda^{N-2a-b}\right), \quad a =
0,...,\left[\frac{\mathrm{N}}{2} \right]\label{Hamiltonians}
\end{eqnarray}
where, by definition, $\mathcal{E}_{a 0}$ is the coefficient of the power
$\lambda^{\mathrm{N}-2a}$.
Among the conserved hamiltonians there are polynomial ones that depend only on the eigenvalues of the Lax operator
and rational ones that instead depend also on the initial twisting data (the $\mathrm{H}^\star$-rotation of the Lax operator from its diagonal form in the case of real eigenvalues or its block-diagonal normal form in the case of
complex eigenvalues). There are also among the $\mathfrak{h}_{ab}$ Casimir functions that have vanishing Poisson brackets with all the canonical variables.
\par
As we extensively discussed in \cite{noiultimo} the classification of spectral types, which corresponds with the classification of normal forms given in \cite{Bergshoeff:2008be}, can be mapped into the foliation of the space of
Lax orbits spanned by the conserved hamiltonians of polynomial type.
\par
For instance in the case of
$\mathrm{SL(3,\mathbb{R})}/\mathrm{SO(1,2)}$, where the general
form of the Lax operator is given by \cite{noiultimo}
\begin{eqnarray}\label{LElax}
    L (t)  =  \left(
\begin{array}{lll}
 \frac{1}{\sqrt{2}}Y_1(t)-\frac{1}{\sqrt{6}}Y_2(t) & -\frac{1}{2}
   Y_3(t) & -\frac{1}{2} Y_5(t) \\
 \frac{1}{2}Y_3(t) &
   -\frac{1}{\sqrt{2}}Y_1(t)-\frac{1}{\sqrt{6}} Y_2(t)& -\frac{1}{2}
   Y_4(t) \\
 \frac{1}{2} Y_5(t)& -\frac{1}{2} Y_4(t) & \sqrt{\frac{2}{3}}
   Y_2(t)
\end{array} \right)
\end{eqnarray}
leading to the following system of differential equations:
\begin{eqnarray}
  -\frac{Y_3(t)^2}{\sqrt{2}}-\frac{Y_4(t)^2}{2
   \sqrt{2}}-\frac{Y_5(t)^2}{2 \sqrt{2}}+\frac{d}{dt} \,Y_1(t) &=& 0~, \nonumber\\
  -\frac{1}{2} \sqrt{\frac{3}{2}} Y_4(t)^2+\frac{1}{2}
   \sqrt{\frac{3}{2}} Y_5(t)^2+\frac{d}{dt} \,Y_2(t) &=& 0~, \nonumber\\
-\sqrt{2} Y_1(t) Y_3(t)-Y_4(t) Y_5(t)+\frac{d}{dt} \,Y_3(t) &=& 0~, \nonumber\\
  \frac{Y_1(t)
   Y_4(t)}{\sqrt{2}}+\sqrt{\frac{3}{2}} Y_2(t)
   Y_4(t)-Y_3(t) Y_5(t)+\frac{d}{dt} \,Y_4(t) &=& 0~, \nonumber\\
  -\frac{Y_1(t)
   Y_5(t)}{\sqrt{2}}+\sqrt{\frac{3}{2}} Y_2(t)
   Y_5(t)+\frac{d}{dt} \,Y_5(t) &=& 0~, \label{diffequeLE}
\end{eqnarray}
there are just a quadratic and a cubic hamiltonian
\begin{eqnarray}
\mathfrak{h}_1  \, \doteq  \, \mathfrak{h}_{01}  &=& 0~,\label{h1Ham} \\
 \mathfrak{h}_2 \, \doteq  \, \mathfrak{h}_{02}   &=& \frac{1}{2} Y_1(t)^2+\frac{1}{2}
   Y_2(t)^2-\frac{1}{4} Y_3(t)^2+\frac{1}{4}
   Y_4(t)^2-\frac{1}{4} Y_5(t)^2~,\label{h2Ham} \\
\mathfrak{h}_3 \, \doteq  \,  \mathfrak{h}_{03}  &=& \frac{Y_2(t)^3}{3 \sqrt{6}}-\frac{Y_1(t)^2
   Y_2(t)}{\sqrt{6}}+\frac{Y_3(t)^2 Y_2(t)}{2
   \sqrt{6}}+\frac{Y_4(t)^2 Y_2(t)}{4
   \sqrt{6}}-\frac{Y_5(t)^2 Y_2(t)}{4
   \sqrt{6}}\nonumber\\
   && -\frac{Y_1(t) Y_4(t)^2}{4
   \sqrt{2}}-\frac{Y_1(t) Y_5(t)^2}{4
   \sqrt{2}}+\frac{1}{4} Y_3(t) Y_4(t) Y_5(t) \label{h3Ham}
   \end{eqnarray}
while the rational hamiltonian is the following one:
   \begin{eqnarray}
\mathfrak{h}_4 \, \doteq  \,  \mathfrak{h}_{11}  &=& \frac{Y_1(t)}{\sqrt{2}}+\frac{Y_2(t)}{\sqrt{6}
   }-\frac{Y_3(t) Y_4(t)}{2 Y_5(t)} \, . \label{h4Ham}
\end{eqnarray}
The space of orbits is separated in two distinct regions by the value of the discriminant
\begin{equation}\label{discriminante}
    \Delta \, \equiv \, -12 \mathfrak{h}_2^3 \, + \, 81 \,
    \mathfrak{h}_3^2 \, .
\end{equation}
\begin{itemize}
  \item In the region where $\Delta < 0$ we have three distinct real eigenvalues.
  \item In the region where $\Delta > 0$ there is one real eigenvalue and a pair of complex conjugate eigenvalues.
  \item The locus $\Delta = 0$ corresponds to orbits admitting an enhanced symmetry, except at the cusp.
\end{itemize}
\iffigs
\begin{figure}
\begin{center}
\includegraphics[height=9cm]{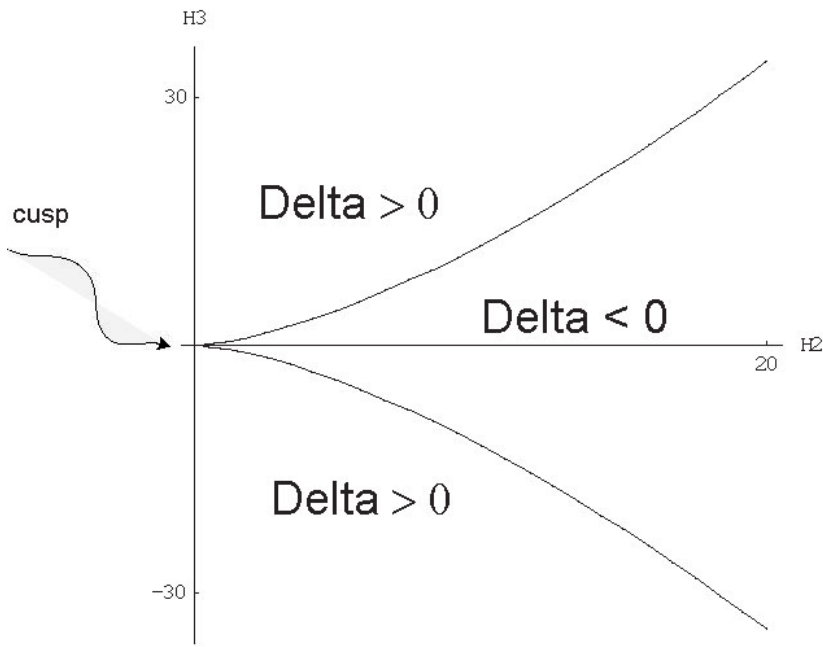}
\end{center}
\else\fi \caption{\label{phasediagr} Orbit structure in the
$\mathfrak{h}_2,\mathfrak{h}_3$ plane. The cuspidal point corresponds to the Nilpotent
orbits}
\end{figure}
\subsection{The orbits of the $\mathrm{J=2}$ representation of $\mathrm{SO(1,2)}$}
The easiest way to discuss the space of orbits for the $\mathrm{SL(3,\mathbb{R})}/\mathrm{SO(1,2)}$ case is to recall that
the coset generators span the five dimensional $\mathrm{J=2}$ representation of $\mathrm{SO(1,2)}$, drawing an analogy to the well known case of orbits of the $\mathrm{J=1}$ representation. In the latter case we deal with vectors
and the space of orbits decomposes into three sectors:
\begin{enumerate}
  \item The orbits of time-like vectors $(v \, ,\,v) \,>0$ which admit $\mathrm{SO(2)}$ as stability subgroup and therefore span the coset $\mathrm{SO(1,2)/SO(2)}$.
  \item The orbits of space-like vectors $(v \, ,\,v) \,< 0$ which admit $\mathrm{SO(1,1)}$ as stability subgroup and therefore span the coset $\mathrm{SO(1,2)/SO(1,1)}$.
\item The orbits of light-like vectors $(v \, ,\,v) \,= 0$ which admit no stability subgroup and lie on the light-cone which separates the two regions of time-like and space-like orbits.
\end{enumerate}
In the $\mathrm{J=2}$ case the role of the light-cone is played by the $\Delta = 0$ locus of vanishing discriminant.
The orbits with $\Delta > 0$ have the eigenvalue structure we have just discussed and admit no stability subgroup. This can be directly verified using the generators of $SO(1,2)$ given in \cite{noiultimo} for the $\mathrm{J=2}$ representation and recalled here for convenience:
\begin{equation}\label{RRJrep}
    \begin{array}{ccccccc}
       \mathrm{R(J_1)}& = & \left(
\begin{array}{lllll}
 ~~0 & 0 & -2 & ~~0 & ~~0 \\
 ~~0 & 0 & ~~0 & ~~0 & ~~0 \\
 -2 & 0 & ~~0 & ~~0 & ~~0 \\
 ~~0 & 0 & ~~0 & ~~0 & -1 \\
 ~~0 & 0 &~~ 0 & -1 &~~ 0
\end{array}
\right) & , & \mathrm{R(J_2)} & = &  \left(
\begin{array}{lllll}
 ~~0 & ~~0 & ~~0 & ~1 & 0 \\
 ~~0 & ~~0 & ~~0 & \sqrt{3} & 0 \\
 ~~0 & ~~0 & ~~0 & ~0 & 1 \\
 -1 & -\sqrt{3} & ~~0 & ~0 & 0 \\
 ~~0 & ~~0 & -1 & ~0 & 0
\end{array}
\right)~, \\
\null & \null & \null & \null &\null & \null &\null \\
       \mathrm{R(J_3)} & = & \left(
\begin{array}{lllll}
~~ 0 &~ 0 & ~~0 & ~~0 & -1 \\
~~ 0 &~ 0 & ~~0 & ~~0 & \sqrt{3} \\
~~ 0 &~ 0 & ~~0 & -1 & ~~0 \\
~~ 0 &~ 0 & -1 & ~~0 & ~~0 \\
 -1 & \sqrt{3} & ~~0 & ~~0 & ~~0
\end{array}
\right) & . & \null & \null & \null
     \end{array}
\end{equation}
In this way the original five dimensional space is parameterized for $\Delta >0$ orbits by the values of the two hamiltonians $\mathfrak{h}_2 \, , \, \mathfrak{h}_3$, which label the orbit, and by the three parameters of $\mathrm{SO(1,2)}$ that span it. Similar conclusion one draws for the $\Delta < 0$. Apart from the difference in spectral type these orbits
behave exactly in the same way. There is no stability subgroup.
\par
On the other hand the locus $\Delta = 0$ contains $5$-vectors that always admit a stabilizer $\mathcal{S}$, either belonging to the compact $\mathcal{S} \in \so(2)$  subalgebra or to a non-compact subalgebra $\mathcal{S} \in \so(1,1)$ of $\so(1,2)$. For these orbits the discriminant $\Delta$ vanishes, yet the individual hamiltonians $\mathfrak{h}_2$ and $\mathfrak{h}_3$ are generically different from zero. In this way  each of the $\Delta=0$ orbits spans either the  $\mathrm{SO(1,2)/SO(2)}$ or the $\mathrm{SO(1,2)/SO(1,1)}$ coset.
\par
There is finally the cuspidal locus displayed in fig.\ref{phasediagr} where both $\mathfrak{h}_2$ and $\mathfrak{h}_3$ are zero. The five-vectors lying in this locus correspond to Nilpotent non-diagonalizable $3 \times 3$ matrices
as, for example, the following:
\begin{equation}\label{example1}
 \Omega \, = \,    \left(
\begin{array}{lll}
 0 & -1 & 0 \\
 1 & ~~0 & 1 \\
 0 & ~~1 & 0
\end{array}
\right) \, .
\end{equation}
This provides an intrinsic characterization of the Nilpotent orbit: vanishing of both polynomial hamiltonians.
\par
Another interesting and equivalent algebraic characterization is the following. The $2$-dimensional space of
non-compact generators of the Lie algebra $\so(1,2)$ given by the span of $\mathrm{J}_1$ and $\mathrm{J}_3$ contains
operators whose eigenvalues are necessarily:
\begin{equation}\label{autovalori}
    -2 \, \ell \, , \, - \ell \, , \, 0 \, , \, +\ell \, , \, +2 \, \ell
\end{equation}
where $\ell$ is an arbitrary normalization of the operator. All the Nilpotent operators $L$ belonging to the singular
cuspidal orbit $\mathcal{C}_0$ are eigenvectors of one of these non-compact generators $\mathrm{G}$ with a finite eigenvalue $\mu \ne 0$:
\begin{equation}\label{cuspidal}
 L \, \in \, \mathcal{C}_0 \, \Leftrightarrow \, \forall \, G \, \in \so(1,2) \,  \, :  \,   \left [G \, , \, L \right ] \, = \, \mu \, L \,.
\end{equation}
On the contrary one can easily verify that no five-vector corresponding to regular orbits can be eigenvector of any of the $\so(1,2)$ generators.
\par
In view of these remarks we see that the $\Delta = 0$ locus is also alternatively characterized as the set of operators that are eigenstates of some $\so(1,2)$ generator. Null eigenstates occur along the two branches and give orbits with enhanced symmetry (stability subgroup). Eigenstates of non-vanishing eigenvalue occur only at the cusp.
\begin{congettura}
For higher groups with more hamiltonians, the secular equation has typically degree higher than four and the discussion
of discriminants becomes unavailable. Yet the second algebraic characterization of cuspidal orbits of Nilpotent operators that,  by definition, have all vanishing eigenvalues and are not diagonalizable, remains viable and appears very promising. Indeed we conjecture that the orbits of such Lax operators can  be found by this method, namely searching for
eigenstates of the non-compact generators of $H^\star$.
\end{congettura}
\par
We stress that in the case of supergravity billiards \cite{sahaedio}, \cite{Fre':2007hd} the isotropy group $\mathrm{H}$ is compact and therefore there are no real Lax operators that can be eigenstates of any generator of $\mathrm{H}$. It is precisely the pseudo-Riemannian nature of
the coset $\mathrm{G/H}^\star$ what allows for the appearance of nilpotent Laxes.
\par
For this singular, isolated class of initial conditions the integration algorithm was not provided in
\cite{noiultimo} since one of its ingredients consists  of the  diagonalization of the initial Lax. The next section fills this gap showing that the solution can be directly and simply constructed in terms of the initial Lax $L_0$.
\section{The integration algorithm for Nilpotent Lax operators}
Let us assume that the initial data for the Lax equation
\begin{eqnarray}
\label{Lax} \frac{d}{dt} L(t)+\left [\,L_>\, -\, L_<\, , \,
L(t)\,\right ] \, = \, 0 \label{newLax}
\end{eqnarray}
are provided by an $\eta$-symmetric matrix:
\begin{equation}\label{etasym}
    L(0) \, = \, L_0 \quad ; \quad L_0 \,\eta \, = \, \eta\, L_0^T
\end{equation}
which is also nilpotent and therefore not diagonalizable. Let us
call $n$ the minimal degree of nilpotency\footnote{In the
$\mathrm{SL(3,\mathbb{R})}$ case the degree of nilpotency is
either $n = 2$ or $n = 3$. It seems that the same occurs also in
the case of more general cosets $\mathrm{G/H}^\star$ occurring in
supergravity theories. However  the algorithm we  present  applies
to any finite degree of nilpotency $n$.}:
\begin{equation}\label{performatre}
    L_0^n \, = \, 0 \, .
\end{equation}
The complete solution for $L(t)$ that admits such an initial condition is constructed in the following way.
First define the following building blocks:
\begin{enumerate}
\item The following $\mathrm{N \times N}$ matrix function:
  \begin{equation}\label{cijN}
    \mathcal{C}(t)\, = e^{-2\, t \, L_0} \,
    = \,  \sum_{k=0}^{n-1} \,
    \frac{1}{k!} \left(-2 t\right)^k \, L_0^{k}\, .
  \end{equation}
  \item The minor functions constructed in the following way:
  \begin{eqnarray}\label{bordoni}
    &&\mathfrak{M}_{ik}(t) =  (-1)^{i+k}\, \mbox{Det} \,
    \left ( \begin{array}{ccc}
    \mathcal{C}_{1,1}(t)  &\dots  & \mathcal{C}_{1,i-1}(t)\\
     \vdots & \vdots &\vdots\\
    \widehat{\mathcal{C}_{k,1}}(t) & \dots & \widehat{\mathcal{C}_{k,i-1}}(t)\\
     \vdots & \vdots & \vdots \\
    \mathcal{C}_{i,1}(t) & \dots &
    \mathcal{C}_{i,i-1}(t)
\end{array} \right ),  \, 1 \leq k \leq i\, ; \,
2\leq  i \leq N\, ,\nonumber\\
&&\mathfrak{M}_{11}(t) =  1
  \end{eqnarray}
  where the hats on the entries corresponding to the $k$-th row mean that such a row has been suppressed
  giving rise to a squared $(i-1) \times (i-1)$ matrix of which one can calculate the determinant.
  \item The determinant functions:
  \begin{equation}\label{DktN}
    \mathfrak{D}_{i}(t) \, = \, \mbox{Det} \, \left ( \begin{array}{ccc}
    \mathcal{C}_{1,1}(t) & \dots & \mathcal{C}_{1,i}(t)\\
    \vdots & \vdots & \vdots \\
    \mathcal{C}_{i,1}(t) & \dots & \mathcal{C}_{i,i}(t)
    \end{array}\right)  \, , \quad
    \mathfrak{D}_{0}(t):=1\, .
    \end{equation}
\end{enumerate}
Then in terms of them we write the entries of the \textit{time}-evolving Lax operator in the following way:
\begin{equation}\label{algorone}
    L_{pq}(t) \, = \, \frac{\eta_{qq}}{\sqrt{\mathfrak{D}_{p}(t)\,\mathfrak{D}_{p-1}(t)\mathfrak{D}_{q}(t)\,\mathfrak{D}_{q-1}(t)}}
    \, \sum_{k=1}^p \,\sum_{\ell=1}^q \, \mathfrak{M}_{pk}(t) \, \left(\mathcal{C}(t)\,L_0\, \eta\right)_{k\ell} \, \mathfrak{M}_{q\ell}(t) \, .
\end{equation}
As an example of the method we give the explicit form of the Lax operator which is produced by the integration
with initial conditions provided by the operator $ \Omega$ displayed in (\ref{example1}):
\begin{equation}\label{solution1}
 L_1(t)  \, = \, \left(
\begin{array}{lll}
 ~\frac{2 t}{1-2 t^2} & -\frac{\sqrt{1+2 t^2}}{1-2t^2} & ~~~0 \\
 \frac{\sqrt{1+2 t^2}}{1-2t^2} &
  ~~~ \frac{4 t}{4 t^4-1} & \frac{\sqrt{1-2 t^2}}{1+2t^2} \\
 ~~~0 & ~~~\frac{\sqrt{1-2 t^2}}{1+2t^2}
   & ~\frac{2 t}{1+2 t^2}
\end{array}
\right) \, .
\end{equation}
This is a representative of the orbit where the Lax operator has degree of nilpotency $n=3$. It corresponds to the eigenvalue $\mu \, = \,1$ of the non compact $\so(1,2)$ generator $\mathrm{R(J_3)}$.
\par
It is quite interesting to employ the algorithm for the square of the matrix $\Omega$ which is an eigenvector corresponding to eigenvalue $\mu = 2$ and has degree of nilpotency $n=2$. The result is a very simple solution with a completely different analytic behaviour
\begin{equation}\label{L2}
   L_2 (t) \, = \,  \frac{1}{1+2t}\, \left(
\begin{array}{lll}
-1 & 0 & -1 \\
~~ 0 & 0 & ~~0 \\
~~ 1 & 0 & ~~1
\end{array}
\right) \, .
\end{equation}
\subsection{Summarizing}
We conclude that the two above presented solutions are the
representatives of the two  nilpotent inequivalent orbits
corresponding to degree of nilpotency $n=3$ and $n=2$,
respectively. Any other solution of the same classes is obtained
by applying our algorithm to an $\mathrm{SO(1,2)}$ rotation of the
corresponding initial data. All in all the three-dimensional
manifold of nilpotent Lax operators is provided by those matrices
(\ref{LElax}) whose five parameters $Y_i$ satisfy the two
constraints $\mathfrak{h}_2(Y)\, = \, \mathfrak{h}_3(Y) \, = \,
0$, according to the definitions (\ref{h2Ham}--\ref{h3Ham}) of the
quadratic and cubic hamiltonians.
\par
The other exceptional orbits are arranged along the two branches of the $\Delta = 0$ locus and each of them spans either an $\mathrm{SO(1,2)/SO(2)}$ or an $\mathrm{SO(1,2)/SO(1,1)}$ coset manifold since there is either an $\so(1,1)$ or an $\so(2)$ stability subalgebra. These orbits correspond to diagonalizable Lax operators and are covered by the algorithm presented in our previous paper \cite{noiultimo}.
\section{Conclusions}
In this note we have shown how the integration algorithm for Lax
equations can be extended also to the case of non-diagonalizable
(see, also the appendix) initial data. Applications to the
construction of extremal BPS solutions is postponed to future
publications.
\appendix
\paragraph{Aknowledgments} We would like to express our gratitude to our frequent collaborator and excellent friend Mario Trigiante for the exchange of useful information we had with him in these last few days. He pointed out to us the
relevance of the nilpotent case and formulated the problem that we were
able to solve in complete generality.
\vskip 0.3cm
\paragraph{Note added in revised version}
We recall that a particular solution corresponding to a specific
nilpotent initial condition  for the
$\mathrm{SL(3,\mathbb{R})}/\mathrm{SO(1,2)}$ case was presented in
the revised version of \cite{marioetal} which appeared on the
hep-th ArXiv the very same day as the first version of the present
paper, where the problem obtained its general solution.
%\newpage
\section{A generalized formula}
Although the physical or geometrical applications of such an
equation are not immediately evident to us we remark that one
could consider a generalization of the Lax equation for symmetric
spaces $\mathrm{G/H}^\star$ by writing the very same differential
condition (\ref{newLax}) imposed on a matrix $L(t)$ which is not
required to satisfy the $\eta$-symmetry conditions (\ref{etasym}).
In other words we can extend Lax equation to matrices $L(t)$ that
are not necessarily elements of the orthogonal complement
$\mathbb{K}$ to a subalgebra $\mathbb{H}^\star \subset \mathbb{G}
\subset \slal (\mathrm{N},\mathbb{R})$.
\par
It is interesting that we can integrate Lax equation in a
completely general way with arbitrary initial Lax matrices $L_0$.
This is done by substituting eq.(\ref{algorone}) with the
following one:
\begin{equation}\label{algorone_general}
    L_{pq}(t) \, = \, \frac{1}{\sqrt{\mathfrak{D}_{p}(t)\,\mathfrak{D}_{p-1}(t)\mathfrak{D}_{q}(t)\,\mathfrak{D}_{q-1}(t)}}
    \, \sum_{k=1}^p \,\sum_{\ell=1}^q \, \mathfrak{M}_{pk}(t) \, \left(\mathcal{C}(t)\, L_0\right)_{k\ell} \, \widetilde{\mathfrak{M}}_{q\ell}(t)
\end{equation}
where
\begin{eqnarray}\label{bordoni_tilde}
    && \widetilde{\mathfrak{M}}_{ik}(t) :=  (-1)^{i+k}\, \mbox{Det} \,
    \left ( \begin{array}{ccccc}
    \mathcal{C}_{1,1}(t)  &\dots & \widehat{\mathcal{C}_{1,k}}(t) &\dots  & \mathcal{C}_{1,i}(t)\\
     \vdots & \vdots & \vdots & \vdots & \vdots\\
 \mathcal{C}_{i-1,1}(t) & \dots & \widehat{\mathcal{C}_{i-1,k}}(t) & \dots &  \mathcal{C}_{i-1,i}(t)
\end{array} \right ),  \nonumber\\
&&\, 1 \leq k \leq i\, \, ; \, \, 2\leq  i \leq \mathrm{N}\quad ;
\quad \widetilde{\mathfrak{M}}_{11}(t) =  1\, ,
  \end{eqnarray}
and all other definitions remaining the same as in the main text.
In eq.(\ref{bordoni_tilde}) the hatted $k$-th column is
  deleted just as in eq.(\ref{bordoni}) it was deleted the $k$-th row.
  \par
This expression gives the solution to Lax equation (\ref{newLax})
for the case of generic initial conditions. It applies in
particular to the case of diagonalizable initial matrices $L_0$.
In that case, however, eq.(\ref{algorone_general}) is only formal
and not too useful since the function $\mathcal{C}(t)$
(\ref{cijN}) is constructed by a matrix exponentiation and
involves summing an infinite series. On the contrary for nilpotent
matrices $L_0$ eq.(\ref{algorone_general}) provides a useful
explicit general integral. The class of nilpotent matrices is much
larger than the class of nilpotent $\eta$-symmetric matrices
(\ref{etasym}), the latter being associated with symmetric spaces.
\par
Although the physical meaning of Lax equation in the more general
context of nilpotent, but not necessarily $\eta$-symmetric
matrices (\ref{etasym}) is so far unknown, yet it is worth
mentioning the existence of this generalized integration formula
whose applications will certainly  be discovered soon.
\par
Now, let us derive another equivalent representation of equations
(\ref{algorone_general}) which will be useful in case of both
nilpotent and non-nilpotent, $\eta$-symmetric and
$\eta$-non-symmetric initial data $L_0$ entering these equations.
\par
At first, we recall that both diagonalizable and
non-diagonalizable (in particular nilpotent) generic initial
conditions for the Lax matrix $L_0$ can be represented in the
Jordan normal form
\begin{equation}\label{Jordan0}
    L_{0} \, = \, \Phi(0)\, \mathcal{J}\, \Phi(0)^{-1}
\end{equation}
where $\Phi(0)$ is an invertible $N\times N$ matrix, $\mathcal{J}$
is a block-diagonal $N\times N$ matrix with the $d_{\alpha}\times
d_{\alpha}$ matrix sub-blocks $J_{\lambda_{\alpha}}$
\begin{eqnarray}\label{Jordan1}
&&\mathcal{J}\,=\,\left(\begin{array}{cccc}
    \mathcal{J}_{\lambda_1} & 0 & \dots & 0\\
    0 & \ddots & \ddots &\vdots \\
    \vdots & \ddots & \ddots &0 \\
    0 & \dots & 0& J_{\lambda_m}
    \end{array}\right)\, ,\quad
    \mathcal{J}_{\lambda_{\alpha}}\,=\,\left(\begin{array}{ccccc}
    \lambda_{\alpha} & 1 & 0 & \dots & 0\\
    0 & \ddots & \ddots &\ddots & \vdots\\
    \vdots & 0 & \ddots & \ddots & 0 \\
      \vdots  & \vdots & \ddots & \ddots & 1 \\
    0 & \dots & \dots & 0 & \lambda_{\alpha}
    \end{array}\right)\, , \\
&& \quad \quad \quad \alpha\,=\, 1,\dots,m\, ,\quad 1\, \leq m \,
\leq N\, ,\quad d_{\alpha}\geq \, 1\, , \quad
\sum_{\alpha=1}^{m}\,d_{\alpha}\,=\,N \nonumber
\end{eqnarray}
and $\lambda_{\alpha}$ are constants. Substituting eq.
(\ref{Jordan0}) into equations (\ref{algorone_general}) and
(\ref{cijN}), the latter can identically be represented in the
following equivalent form:
\begin{eqnarray}\label{Laxgennew}
  &&  L(t)\, = \,\Phi(t)\, \mathcal{J}\, \Phi(t)^{-1}\, ,\\
&& \mathcal{C}(t)\, = \,\Phi(0)\, e^{-2t\,\mathcal{J}}\,
\Phi(0)^{-1}\, \label{Cnew}
  \end{eqnarray}
where
\begin{eqnarray}
\Phi_{ij}(t)&=&\frac{1}{\sqrt{\mathfrak{D}_i(t)\mathfrak{D}_{i-1}(t)}}
\,\sum_{s=1}^{N}\, \mathrm{Det} \, \left(\begin{array}{cccc}
\mathcal{C}_{1,1}(t)&\dots &\mathcal{C}_{1,i-1}(t)& \Phi_{1s}(0)\\
\vdots&\vdots&\vdots&\vdots\\
\mathcal{C}_{i,1}(t)&\dots &\mathcal{C}_{i,i-1}(t)& \Phi_{is}(0)\\
\end{array}\right)\,\left(e^{-t\,\mathcal{J}}\right)_{sj},\nn\\
\left(\Phi^{-1}\right)_{ji}(t)&=&\frac{1}{\sqrt{\mathfrak{D}_i(t)\mathfrak{D}_{i-1}(t)}}
\,\sum_{s=1}^{N}\, \left(e^{-t\,\mathcal{J}}\right)_{js}
\mathrm{Det} \, \left (
\begin{array}{ccc}
\mathcal{C}_{1,1}(t)&\dots &\mathcal{C}_{1,i}(t)\\
\vdots&\vdots&\vdots\\
\mathcal{C}_{i-1,1}(t)&\dots &\mathcal{C}_{i-1,i}(t)\\
\left(\Phi^{-1}\right)_{s1}(0)&\dots &
\left(\Phi^{-1}\right)_{si}(0)
\end{array}\right
)\nonumber\\ \label{Gensolut}
 \end{eqnarray}
and the functions $\mathfrak{D}_i(t)$ are defined in eq.
(\ref{DktN}).
\par
For the particular case of diagonalizable initial conditions, i.e.
when $m\,=\,N$ ($d_{\alpha}\,=\,1$, $\alpha\,=\, 1,\dots,N$), the
corresponding matrix $\mathcal{J}$ (\ref{Jordan1}) becomes a
diagonal matrix of the $N$-eigenvalues $\lambda_{\alpha}$
$(\alpha\,=\, 1,\dots,N)$, and equations
(\ref{Laxgennew}--\ref{Gensolut}) reproduce the general solutions
to Lax equations (\ref{newLax}) for the case of {\it generic
diagonalizable} Lax matrices derived in \cite{kodama2} for the
first time. What concerns expressions
(\ref{Laxgennew}--\ref{Gensolut}) in the case $m\, \neq \, N$,
they are a generalization of the above-mentioned general solutions
to the case of {\it generic non-diagonalizable} Lax matrices
constructed, to our best knowledge, for the first time in the
present paper.
\par
When deriving eqs. (\ref{Laxgennew}--\ref{Gensolut}) starting from
eq. (\ref{algorone_general}), we did not use any properties of the
Lax operators $L_0$ but its rather general representation
(\ref{Jordan0}). In case of diagonal matrices $\mathcal{J}$ we
proved the correctness of derived expressions
(\ref{Laxgennew}--\ref{Gensolut}) by their established
relationship with the earlier known solution \cite{kodama2} of Lax
equations (\ref{newLax}), as it was already mentioned above. We
also verified their correctness in case of
$\frac{SL(N)}{SO(p,N-p)}$ cosets at $N=2\,,3\,,4$ and $5$ for
non-diagonalizable nilpotent matrices $\mathcal{J}$. Altogether,
this gives an evidence in favour of correctness of eqs.
(\ref{Laxgennew}--\ref{Gensolut}) in case of generic matrices
$\mathcal{J}$ (\ref{Jordan0}) as well.
\par
The constructed general solutions
(\ref{Laxgennew}--\ref{Gensolut}) are explicitly parameterized by
the initial data encoded in the matrices $\mathcal{J}$ and
$\Phi(0)$. In order these solutions could be expressed in terms of
elementary functions, the exponential $e^{-t\,\mathcal{J}}$,
entering into relations (\ref{Cnew}) and(\ref{Gensolut}), has to
admit a closed form in terms of elementary functions. Obviously,
it is a very simple task for diagonal and nilpotent matrices
$\mathcal{J}$. Thus, in the latter case all eigenvalues are zero
since $\mathcal{J}$ in (\ref{Jordan1}) is a nilpotent matrix, and
the expansion of the exponential $e^{-t\,\mathcal{J}}$ in power
series with respect to $\mathcal{J}$ terminates at some finite
order. It is more complicated but in principle a solvable task for
many interesting, more generic non-nilpotent and
non-diagonalizable cases.
\par
Thus, for an example in case of $\eta$-symmetric Lax operators
$L_0$ (\ref{etasym}), $\Phi(t)$ is a pseudo-orthogonal matrix,
$\Phi(t)\,\in \, SO(p,q)$, i.e. $\Phi^{-1}(t)\,=\,\eta\,
\Phi^{T}(t)\,\eta$. In this case the non-diagonalizable matrices
$\mathcal{J}$ have a very simple structure
\cite{Bergshoeff:2008be}:
\begin{equation}\label{Jordanetasymmetric}
    \mathcal{J}\,=\,Q\,+\,Nil\, , \quad [Q\,,\,Nil]\,=\,0
\end{equation}
where $Nil$ is a nilpotent matrix and $Q$ is a block-diagonal
matrix with $2\times 2$ and $1\times 1$ sub-blocks
\cite{Bergshoeff:2008be,noiultimo}. Therefore, the exponential
$e^{-t\,\mathcal{J}}$ can be factorized
\begin{equation}\label{factorization}
   e^{-t\,\mathcal{J}}\,=\,e^{Q}\,\,e^{Nil}
\end{equation}
and both exponentials, entering on the r.h.s. of the latter
relation, can easily be expressed in terms of elementary
functions. A detailed discussion of this interesting case will be
presented elsewhere.

\end{document}